\documentclass[fleqn,usenatbib]{mnras}

\usepackage{newtxtext,newtxmath}
\usepackage{hyperref}

\usepackage[T1]{fontenc}
\usepackage{ae,aecompl}
\usepackage{epstopdf}

\usepackage{graphicx}	
\usepackage{amsmath}	
\usepackage{amssymb}	

\newcommand{\Imcyg}{\mbox{$I^2_{m,\mathrm{Cyg}}$}}
\newcommand{\Itcyg}{\mbox{$I^2_{t,\mathrm{Cyg}}$}}
\newcommand{\Imcas}{\mbox{$I^2_{m,\mathrm{Cas}}$}}
\newcommand{\Itcas}{\mbox{$I^2_{t,\mathrm{Cas}}$}}
\newcommand{\Fmcyg}{\mbox{$F_{m, \mathrm{Cyg}}$}}
\newcommand{\Fmcas}{\mbox{$F_{m, \mathrm{Cas}}$}}
\newcommand{\Ftcyg}{\mbox{$F_{t, \mathrm{Cyg}}$}}
\newcommand{\Ftcas}{\mbox{$F_{t, \mathrm{Cas}}$}}

\title[Fading of Cassiopeia A]{The fading of Cassiopeia~A, and improved models for the absolute spectrum of primary radio calibration sources}

\author[Trotter et al.]
{A.~S.~Trotter$^{1}$, 
D.~E.~Reichart$^{1}$\thanks{E-mail: reichart@unc.edu (DER)}, 
R.~E.~Egger$^{1}$, 
J.~St\'{y}blov\'{a}$^{1}$, 
M.~L.~Paggen$^{1}$, \newauthor
J.~R.~Martin$^{1}$, 
D.~A.~Dutton$^{1}$,
J.~E.~Reichart$^{2}$,
N.~D.~Kumar$^{1}$, 
M.~P.~Maples$^{1}$, \newauthor
B.~N.~Barlow$^{1,3}$,
T.~A.~Berger$^{1,4}$, 
A.~C.~Foster$^{1}$, 
N.~R.~Frank$^{1}$,
F.~D.~Ghigo$^{5}$, \newauthor
J.~B.~Haislip$^{1}$, 
S.~A.~Heatherly$^{5}$, 
V.~V.~Kouprianov$^{1}$, 
A.~P.~LaCluyz\'{e}$^{1,6}$,  
D.~A.~Moffett$^{7}$, \newauthor
J.~P.~Moore$^{1}$,
J.~L.~Stanley$^{8}$, 
S.~White$^{5}$
\\
$^{1}$Department of Physics and Astronomy, University of North Carolina at Chapel Hill, Chapel Hill, NC 27599
\\
$^{2}$Jordan High School, Durham, NC 27707
\\
$^{3}$Department of Physics, High Point University, High Point, NC 27268
\\
$^{4}$Institute for Astronomy, University of Hawaii, Honolulu, HI 96822
\\
$^{5}$Green Bank Observatory, Green Bank, WV, 24944
\\
$^{6}$Department of Physics, Central Michigan University, Mount Pleasant, MI 48859
\\
$^{7}$Department of Physics, Furman University, Greenville, SC, 29613
\\
$^{8}$Michigan State University, Lansing, MI, 48824}

\begin{document}

\date{Accepted 2017 March 29. Received 2017 March 29; in original form 2017 March 20}

\pubyear{2017}


\maketitle


\begin{abstract}
Based on five years of observations with the 40-foot telescope at Green Bank Observatory (GBO), \citet{r00} found that the radio source Cassiopeia~A had either faded more slowly between the mid-1970s and late 1990s than \citet{b77} had found it to be fading between the late 1940s and mid-1970s, or that it had rebrightened and then resumed fading sometime between the mid-1970s and mid-1990s, in L band (1.4~GHz).  Here, we present 15 additional years of observations of Cas~A and Cyg~A with the 40-foot in L band, and three and a half additional years of observations of Cas~A, Cyg~A, Tau~A, and Vir~A with GBO's recently refurbished 20-meter telescope in L and X (9~GHz) bands.  We also present a more sophisticated analysis of the 40-foot data, and a reanalysis of the \citet{b77} data, which reveals small, but non-negligible differences.  We find that overall, between the late 1950s and late 2010s, Cas~A faded at an average rate of $0.670 \pm 0.019$~\%/yr in L band, consistent with \citet{r00}.  However, we also find, at the 6.3$\sigma$ credible level, that it did not fade at a constant rate.  Rather, Cas~A faded at a faster rate through at least the late 1960s, rebrightened (or at least faded at a much slower rate), and then resumed fading at a similarly fast rate by, at most, the late 1990s.  Given these differences from the original \citet{b77} analysis, and given the importance of their fitted spectral and temporal models for flux-density calibration in radio astronomy, we update and improve on these models for all four of these radio sources.  In doing so, we additionally find that Tau~A is fading at a rate of $0.102^{+0.042}_{-0.043}$~\%/yr in L band.
\end{abstract}


\section{Introduction}\label{intro}

\begin{table*}
\caption{Summary of GBO telescopes and bands.}\label{tabtel}
\centering
\begin{tabular}{c c c c c}
\hline 
\hline 
Telescope & Band & Central Frequency & Bandwidth & Beamwidth \\
\hline 
\hline 
40-foot  &  L  &  1405~MHz  &  110~MHz  & $1.2^\circ$	\\
20-meter  &  L  &  1395~MHz  &  80~MHz  & $0.75^\circ$ \\
20-meter  &  X  &  9000~MHz  &  2000~MHz  & $0.12^\circ$ \\
\hline 
\end{tabular}
\end{table*}

The bright, but fading, supernova remnant Cassiopeia~A serves as a primary calibration source for radio astronomy (e.g., \citealt{b77}).  As such, it is important that its absolute spectrum and frequency-dependent fading rate be known as accurately as possible.  To this end, \citet{b77} compiled:  (1) absolute measurements of the flux density of Cas~A, made between 1959 and 1973, and at frequencies ranging from 10~MHz to 22~GHz; and (2) measurements of the fading rate of Cas~A, made over 8 -- 20 year spans between 1949 and 1976, and at frequencies ranging from 82~MHz to 9.4~GHz.  To the former data set, they fit their spectral model (see \S\ref{ball}).  To the latter data set, they fit their temporal model, and found that Cas~A was fading at a frequency-dependent rate of (also, see \citealt{d74}):
\begin{equation}\label{eq1}
-\frac{100}{F_\nu}\frac{dF_\nu}{dt} = 0.97(\pm 0.04)-0.30(\pm 0.04)\log\nu_{\mathrm{GHz}}~\mbox{\%/yr} \, .
\end{equation}
\noindent
This corresponds to a fading rate of $0.93 \pm 0.04$~\%/yr at the L-band frequency $\nu = 1405$~MHz.

However, \citet{r00} made additional 1405-MHz measurements of Cas~A, relative to Cyg~A, between 1995 and 1999 with the 40-foot telescope at Green Bank Observatory (GBO) in West Virginia, and found a different result:  In combination with the fitted Cas~A and Cyg~A spectral models of \citet{b77}, corresponding to an epoch of 1965 and evaluated at the same L-band frequency, \citet{r00} found a fading rate of only $0.62 \pm 0.12$~\%/yr.  

In this paper, we present 18 more years of GBO observations, for a total of:  (1) 20 years of 40-foot observations of Cas~A and Cyg~A in L band; and (2) three and a half years of observations with GBO's recently refurbished 20-meter telescope (Martin et al., in prep.), of Cas~A, Cyg~A, Tau~A, and Vir~A in L and X bands (see Table~\ref{tabtel} for telescope and band information).  We present these GBO observations, and review the \citet{b77} data, in \S\ref{data}.  In \S\ref{gbomodel}, we:  (1) present a more sophisticated analysis of the 40-foot data, accounting for small pointing errors and focus differences; (2) measure the fading rate of Cas~A between 1994 and 2017 in L band; and (3) confirm the discrepancy with \citet{b77} identified by \cite{r00}.  In \S\ref{baars}, we reanalyze the \citet{b77} data, and find somewhat different results, but still the discrepancy.  In \S\ref{combine}, we combine the \citet{b77} data, which also include absolute measurements of Cyg~A and Tau~A, and both absolute and relative measurements of Vir~A, with our GBO data, and present new, and simultaneously fitted, spectral and temporal models for all four of these sources, so they may better serve as calibration sources in radio astronomy.  We summarize our conclusions in \S\ref{conclusion}.

\section{Data}\label{data}

\subsection{Baars et al.\@ (1977)}\label{historical}

In addition to measurements of the fading rate of Cas~A (\S\ref{intro}), \citet{b77} compiled absolute measurements of the flux density of Cas~A, Cyg~A, and Tau~A, and both absolute and relative measurements of the flux density of Vir~A, at frequencies ranging from 10~MHz to 35~GHz, from multiple papers published between 1960 and 1975.  These data were collected using a variety of instruments, including dipole, horn, and dish antennae.  They assigned each measurement a weight that was inversely proportional to the square of its published uncertainty.  However, for dish antennae, they decreased this weight by a factor of four, given concerns over how accurately dish antennae could be absolutely calibrated.  In their spectral fits, they included only (1) measurements with less than $6\%$ uncertainty, and (2) measurements between 22~MHz and 22~GHz. 

\begin{table}
\caption{Twenty years of GBO 40-foot L-band measurements of:  (1) the Cas~A-to-Cyg~A flux-density ratio, $R_m\equiv \Fmcas\Imcas/\Fmcyg\Imcyg$ (see \S\ref{40ftresults}); (2) our measurements of the square of the integral of the corresponding peak-normalized source functions, $\Imcas$ and $\Imcyg$; and (3) the weight that we have assigned to each ratio (\S\ref{gbo40data}).}\label{tab1}
\centering
\begin{tabular}{c c c c c}
\hline 
\hline 
Epoch	&	$R_m$	&	$\Imcas$	&	$\Imcyg$	&	Weight	\\
\hline 
\hline 
1995.62	&	1.316	&	1.474	&	1.435	&	1	\\
1996.62	&	1.335	&	1.407	&	1.360	&	1	\\
1996.62	&	1.333	&	1.394	&	1.360	&	1	\\
1997.62	&	1.335	&	1.394	&	1.340	&	1	\\
1999.62	&	1.244	&	1.395	&	1.398	&	1	\\
2000.62	&	1.256	&	1.394	&	1.394	&	1	\\
2000.62	&	1.257	&	1.395	&	1.394	&	1	\\
2002.62	&	1.246	&	1.472	&	1.455	&	1	\\
2002.62	&	1.260	&	1.492	&	1.455	&	1	\\
2003.59	&	1.181	&	1.484	&	1.450	&	1	\\
2004.61	&	1.213	&	1.485	&	1.454	&	 2/3	\\
2004.61	&	1.200	&	1.460	&	1.454	&	 2/3	\\
2004.61	&	1.196	&	1.476	&	1.454	&	 2/3	\\
2004.61	&	1.267	&	1.485	&	1.439	&	 2/3	\\
2004.61	&	1.253	&	1.460	&	1.439	&	 2/3	\\
2004.61	&	1.249	&	1.476	&	1.439	&	 2/3	\\
2006.53	&	1.248	&	1.462	&	1.425	&	1	\\
2006.53	&	1.245	&	1.468	&	1.425	&	1	\\
2007.45	&	1.167	&	1.418	&	1.404	&	 2/3	\\
2007.45	&	1.170	&	1.442	&	1.404	&	 2/3	\\
2007.45	&	1.183	&	1.413	&	1.404	&	 2/3	\\
2007.45	&	1.169	&	1.418	&	1.415	&	 2/3	\\
2007.45	&	1.172	&	1.442	&	1.415	&	 2/3	\\
2007.45	&	1.185	&	1.413	&	1.415	&	 2/3	\\
2009.57	&	1.243	&	1.450	&	1.391	&	1	\\
2009.57	&	1.219	&	1.423	&	1.391	&	1	\\
2010.49	&	1.133	&	1.435	&	1.436	&	1	\\
2010.49	&	1.220	&	1.490	&	1.436	&	1	\\
2010.49	&	1.200	&	1.510	&	1.436	&	1	\\
2011.47	&	1.156	&	1.497	&	1.456	&	1	\\
2011.47	&	1.106	&	1.461	&	1.456	&	1	\\
2011.47	&	1.108	&	1.472	&	1.456	&	1	\\
2011.47	&	1.164	&	1.496	&	1.456	&	1	\\
2011.47	&	1.112	&	1.453	&	1.456	&	1	\\
2012.54	&	1.106	&	1.454	&	1.466	&	 2/3	\\
2012.54	&	1.018	&	1.422	&	1.466	&	 2/3	\\
2012.54	&	1.108	&	1.455	&	1.466	&	 2/3	\\
2012.54	&	1.122	&	1.454	&	1.451	&	 2/3	\\
2012.54	&	1.032	&	1.422	&	1.451	&	 2/3	\\
2012.54	&	1.123	&	1.455	&	1.451	&	 2/3	\\
2013.51	&	1.032	&	1.427	&	1.503	&	1	\\
2013.51	&	1.050	&	1.516	&	1.503	&	1	\\
2013.51	&	0.993	&	1.396	&	1.503	&	1	\\
2013.51	&	1.039	&	1.473	&	1.503	&	1	\\
2014.43	&	1.153	&	1.446	&	1.428	&	1	\\
2014.43	&	1.137	&	1.464	&	1.428	&	1	\\
2014.43	&	1.152	&	1.449	&	1.428	&	1	\\
\hline 
\end{tabular}
\end{table}

In our own fits, we include the same data that \citet{b77} included in their fits (with a few, relatively minor exceptions\footnote{In their Cyg~A and Tau~A spectral fits, \citet{b77} included a few additional, relative measurements from \citet{b72} and \citet{b65}, but these are not clearly specified so we did not include them in our fits.  Additionally, in their Table~4, \citet{b77} averaged three relative measurements of Vir~A from \citet{b72} and, separately, three absolute measurements of Vir~A from \citet{j74}.  We have replaced these with their original, individual measurements.}), and with the same weights.  However, before carrying out their Cas~A spectral fits, \citet{b77} first corrected, or referenced, their data to epoch 1965, using their fitted temporal model (Equation \ref{eq1}).  We on the other hand do not reference their, or our, data to a common epoch; rather, we fit our spectral and temporal models simultaneously (see \S\ref{ball}, \S\ref{combine}).

\subsection{GBO 40-foot}\label{gbo40data}

From 1994 through 2015, we observed Cas~A and Cyg~A in L band (centered at 1405~MHz) using GBO's 40-foot telescope.  The 40-foot is a transit telescope, so we observed each source no more than once per day, as it transited the meridian.  Each year, observations were carried out over a single week, resulting in 1 -- 5 successful Cas~A observations and 0 -- 2 successful Cyg~A observations.  The telescope is controlled manually, and observations were carried out in drift-scan mode, in the following way.  Well before transit, we would move the telescope to the approximate transit elevation of the source, and then begin moving the telescope back and forth in elevation, (1) searching for the source, and (2) collecting background information at different elevations.\footnote{Due to terrestrial spillover, the background level is sensitive to the angle that the telescope makes with respect to the ground.  Since the final elevation is not known at the beginning of the observation, we were careful to collect background information at all possible final elevations.}  Output powers were saved digitally, but were also displayed in real time using an analog strip-chart recorder.  Once the source was detected, we narrowed our elevation range, until, using the strip-chart recorder, we found the elevation for which the signal was greatest.  Experienced operators had enough time to check this multiple times before transit, and were likely able to point the telescope to within a tenth of a beam (i.e., $\approx$$0.1\times1.2^\circ$) of the source's true transit elevation.

Once this was achieved, and prior to the transit of the source, the telescope's elevation was locked, to prevent the telescope from slowly drifting off of the source.  The observation would then continue until the background level was again achieved, and maintained for a few minutes.  Both before and after the observation, a noise diode in the receiver was activated for one minute, and then deactivated for one minute, permitting the observation to be gain calibrated, which we did by linearly interpolating between the two, albeit nearly identical, calibrations.  In post-processing, we excised data that were not collected near the final, locked-in elevation.

\begin{figure*}
 \begin{center}
 \includegraphics[width=\textwidth]{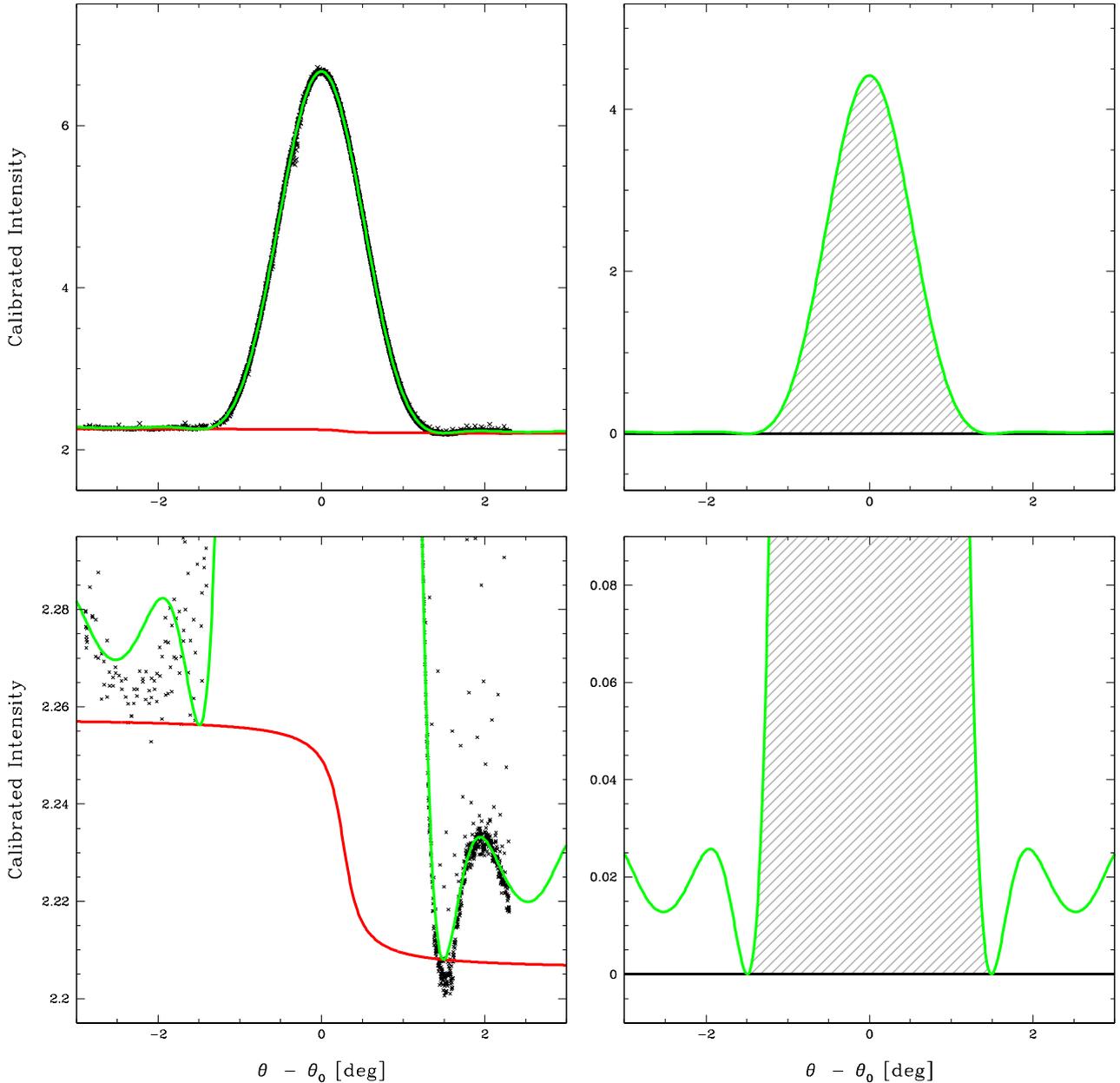}
 \caption{Example drift scan of Cas~A, acquired with the 40-foot in L band. \textbf{Top left:}  Measured intensity, in gain-calibration units, vs.\@ angular position (black crosses), and best-fit source (green curve) plus background (red curve) functions.  Data not collected near the final, locked-in elevation have already been excised.  \textbf{Bottom left:}  An expanded view, to better see the base of the best-fit source function and the best-fit background function.  \textbf{Right:}  Best-fit source function only (i.e., the background-subtracted best-fit model).  The shaded area, divided by the source function's peak value, $F_m$, defines $I_m$ (Equation~\ref{eq_Im}).}
 \label{fig_prof} 
 \end{center}
\end{figure*}

We then modeled each drift scan, $F(\theta)$, as the sum of two functions: a source function, $F_s(\theta)$, and a background function, $F_b(\theta)$ (see Figure~\ref{fig_prof}).  We took the source function to be a non-negative, even function of the form:
\begin{equation}\label{eq-profile}
F_s(\theta) = \left\{{\sum_{n=1}^{N}{a_{n}\cos\left[2\pi n\frac{\left(\theta-\theta_0\right)}{w_\theta}\right]}}\right\}^2 \, .
\end{equation}
The angle at which this function peaks, $\theta_0$, the parameter that determines its width, $w_\theta$, and the $N$ coefficients, $a_n$, were fitted to each drift scan.  We found that $N=6$ was sufficient in each case.

The background function, $F_b(\theta)$, models emission from the Milky Way and terrestrial spillover, as well as any signal drift over the course of the observation, characteristic of the instrument's stability (generally, a very small effect).  We took the background function to be odd, to prevent degeneracies with the source-function parameters:
\begin{equation}\label{eq-background}
F_b(\theta) = b_0 + b_1\tan^{-1}\left(\frac{\theta-b_2}{b_3}\right) \, ,
\end{equation}
where $b_0$, $b_1$, $b_2$, and $b_3$ were also fitted to each drift scan, simultaneously with the source-function parameters.\footnote{For all of the fits in this paper, we used \textit{Galapagos}, which is in-house, parallelized, genetic algorithm-based model-fitting software.  It is particularly good at fitting large-dimensional models (in the above case, 12-dimensional) without being waylaid by local solutions.}  This background function is sufficiently flexible to model most likely behaviors, including everything from constant, linear, and some non-linear functions to step functions (corresponding to a sudden jump or drop in the signal level).

\begin{figure*}
\begin{center}
\includegraphics[width=0.9\textwidth]{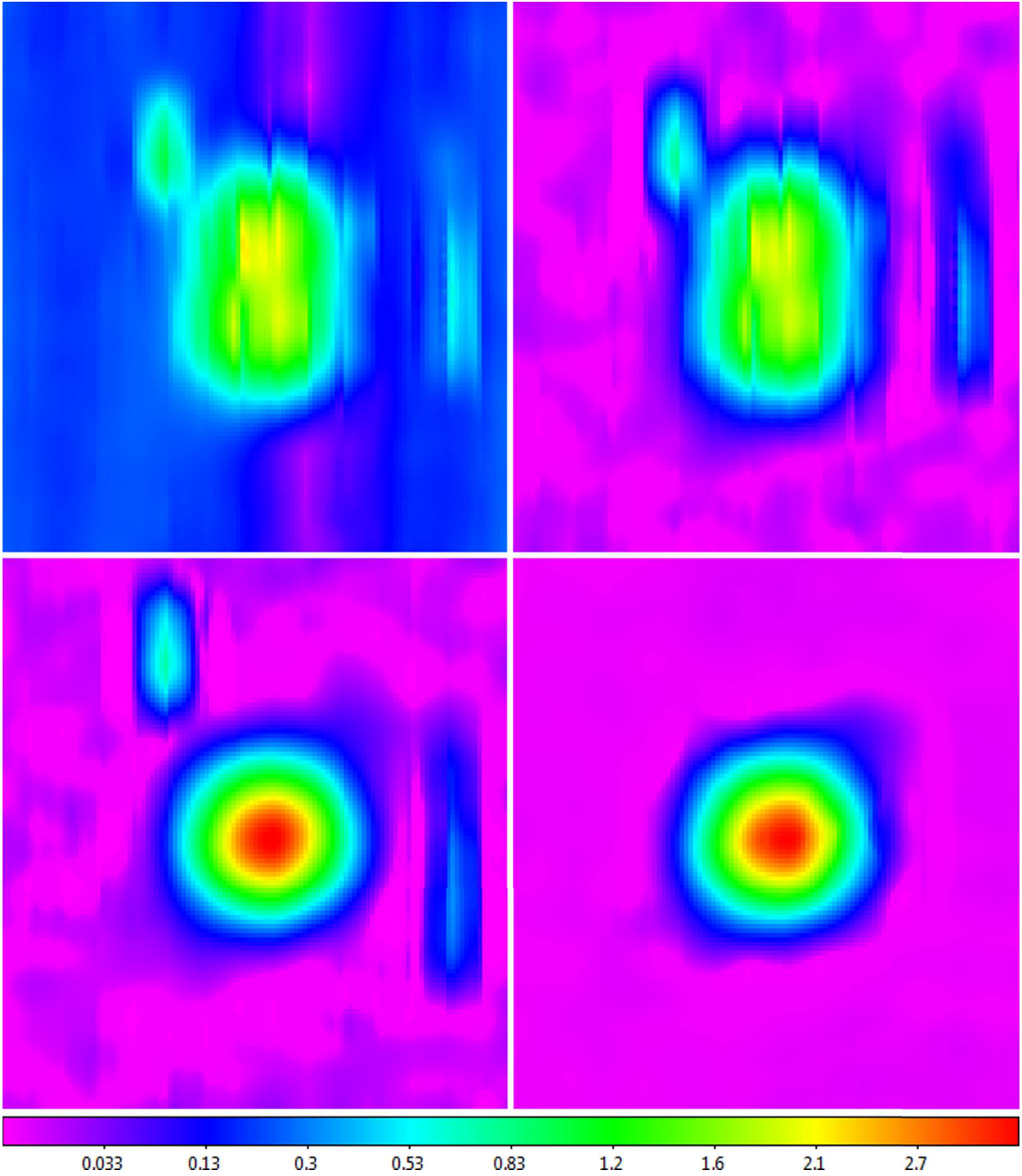}
\caption{Example on-the-fly raster of Cas~A, acquired with the 20-meter in X band.  For all of our 20-meter observations, signal was integrated and read out every 0.2 beams (corresponding to every $7\arcmin$ in this, X-band case) along scans in declination (plotted here vertically), and scans were separated also by 0.2 beams, in right ascension (plotted here horizontally).  In this case, we have zoomed into the central 5.7 beams $\times$ 5.7 beams, corresponding to $40\arcmin$ $\times$ $40\arcmin$.  \textbf{Top left:}  Gain-calibrated measurements.  \textbf{Top right}:  After background subtraction on the recommended, 6-beam scale.  \textbf{Bottom left}:  After time-delay correction, by which we mean that occasionally the clocks of the telescope's separate coordinate and signal computers lose synchronization, which results in a hysteretic splitting of the source along the scan direction, prior to correction.  \textbf{Bottom right}:  After RFI cleaning, on the recommended, 0.8-beam scale.  In all four panels, scale-weighted surface models have been fitted to the data about each pixel, for visualization.  In the first three panels, a minimal scale is used, for maximal surface flexibility.  In the fourth panel, the recommended, 1/3-beam (near-Nyquist) scale is used, and this is the final image.  A non-linear color spectrum is used to emphasize fainter structures (units are dimensionless, with one corresponding to the noise diode).  Details regarding the Skynet Robotic Telescope Network's single-dish image-reduction pipeline will be available in Martin et al.\@ (in prep.).}
\label{fig2}
\end{center}
\end{figure*} 

\begin{figure*}
\begin{center}
\includegraphics[width=0.9\textwidth]{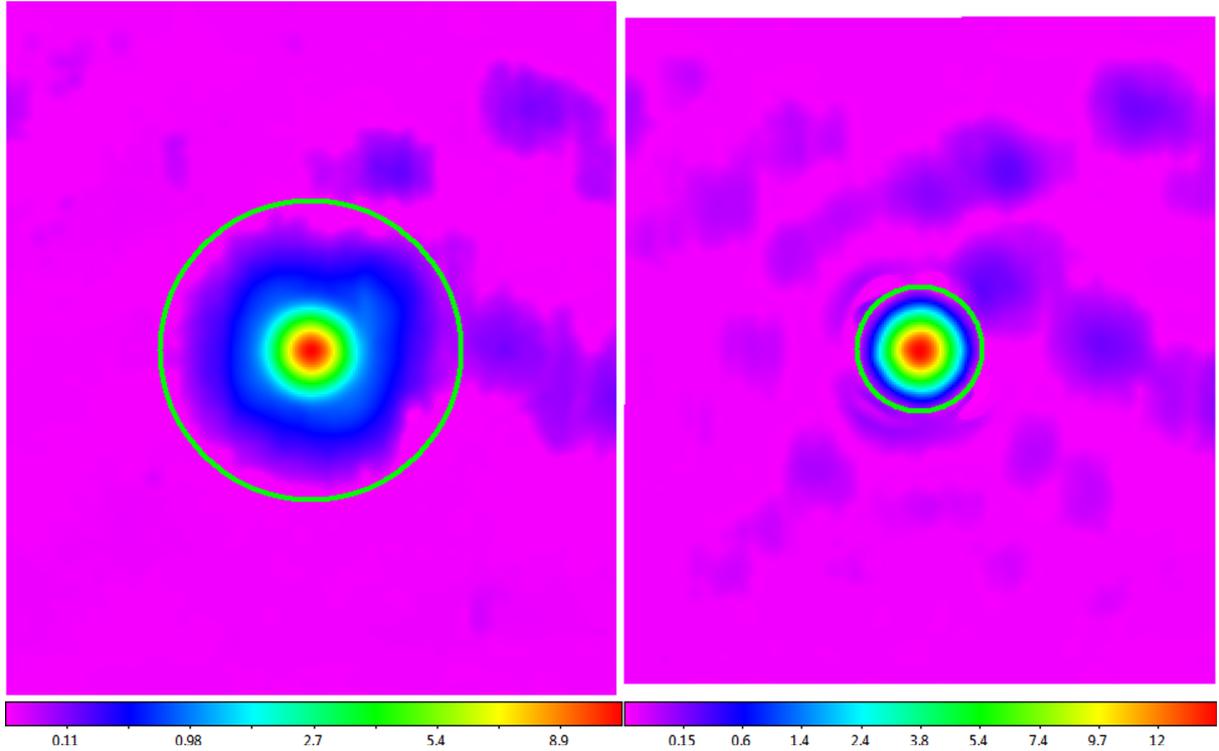}
\caption{Example L-band images of Cas~A, made with the 20-meter in 2014 (left) and 2016 (right).  Each is 13.5 beams $\times$ 13.5 beams, corresponding to $10.1^\circ$ $\times$ $10.1^\circ$, with declination plotted vertically and right ascension plotted horizontally.  A non-linear color spectrum is used to emphasize fainter structures (units are dimensionless, with one corresponding to the noise diode).  The 2014 images are somewhat out of focus, requiring a larger aperture (green circle) for photometry.  This had been corrected by 2016.}
\label{fig3}
\end{center}
\end{figure*}

From each best fit, we calculated:  (1) the measured peak intensity of the source:
\begin{equation}
F_m\equiv F_s(\theta_0) = \left(\sum_{n=1}^{N}{a_n}\right)^2 \, ;
\end{equation}
\noindent
and (2) the integral of the measured source function, normalized to its peak value:
\begin{equation}\label{eq_Im}
I_m \equiv \frac{1}{F_m}\int_{\theta_0-\Delta\theta}^{\theta_0+\Delta\theta}{F_s(\theta)d\theta} \, ,
\end{equation}
where the integration limits are the symmetric, first minima on either side of the source function (Figure~\ref{fig_prof}, bottom right).  In \S\ref{gbomodel}, we show that:  (1) the Cas~A-to-Cyg~A flux-density ratio, $R_m$, depends not only on $\Fmcas$ and $\Fmcyg$, but also on $\Imcas$ and $\Imcyg$; and (2) $\Imcas$ and $\Imcyg$ can also be used to account for small errors in pointing and differences in focus, at least in part.

We list these ratios, $R_m$, and their corresponding values of $\Imcas$ and $\Imcyg$, in Table~\ref{tab1}, for all same-week combinations of our Cas~A and Cyg~A measurements.  For most years, we have only one Cyg~A measurement, having made other observations during Cyg~A's transit time the rest of the week.  For years where we have two Cyg~A measurements, and three Cas~A measurements, we assigned each ratio a weight of $2/3$, given redundancy of information in the six possible combinations.  In total, we obtained 47 (41 non-redundant) L-band measurements of $R_m$ with the 40-foot over a 20-year period.

\begin{table}
\caption{Summary of GBO 20-meter L- and X-band observations.}\label{tab2}
\centering
\begin{tabular}{c c c c}
\hline 
\hline 
Dates & Band & Source & Number of Observations \\
\hline 
\hline 
10 -- 14 Jul 2014 &  L  &  Cas~A  &  24	\\
10 -- 14 Jul 2014 &  L  &  Cyg~A  &  24 \\
17 Jul -- 5 Aug 2014 &  X  &  Cas~A  &  31	\\
17 Jul -- 5 Aug 2014 &  X  &  Cyg~A  &  30 \\
24 -- 29 Sep 2016 &  L  &  Cas~A  &  20	\\
24 -- 29 Sep 2016 &  L  &  Cyg~A  &  23 \\
6 -- 8 Jan 2017 &  L  &  Cas~A  &  23	\\
5 -- 8 Jan 2017 &  L  &  Cyg~A  &  25 \\
5 -- 8 Jan 2017 &  L  &  Tau~A  &  23	\\
5 -- 8 Jan 2017 &  L  &  Vir~A  &  24 \\
\hline 
\end{tabular}
\end{table}

\subsection{GBO 20-meter}\label{gbo20data}

In 2014, 2016, and 2017, we observed different combinations of Cas~A, Cyg~A, Tau~A, and Vir~A in different combinations of L (centered at 1395~MHz) and X (centered at 9000~MHz) bands using GBO's 20-meter telescope (see Table~\ref{tab2}).  The 20-meter had recently been refurbished and integrated into the Skynet Robotic Telescope Network (Martin et al., in prep.).  We submitted each observation through Skynet's web-based user interface,\footnote{https://skynet.unc.edu} after which they were automatically scheduled and completed.  Each L-band observation consisted of a 13.5-beam $\times$ 13.5-beam on-the-fly raster, with 0.2-beam spacing between measurements, both along and between scans, and at a telescope slew speed of 0.5$^\circ$/s took 23.0 minutes to complete.  Each X-band observation consisted of an 18-beam $\times$ 18-beam on-the-fly raster, also with 0.2-beam spacing between measurements, and at a telescope slew speed of 0.3$^\circ$/s took 10.5 minutes to complete.  In total, 247 observations were completed.

Each observation was gain calibrated, background subtracted, time-delay corrected, RFI cleaned, and surface modeled using Skynet's single-dish radio telescope image-reduction pipeline (Martin et al., in prep.; see Figure~\ref{fig2}).  We then photometered each image using 3- and 1.5-beam diameter apertures for the 2014 L- and X-band images, respectively, and using 1.25-beam diameter apertures for the 2016 and 2017 L-band images (focus had been improved for the later L-band images; see Figure~\ref{fig3}).

We determined flux-density ratios by pairing each Cas~A, Tau~A, or Vir~A measurement with its closest-in-time Cyg~A measurement, and by pairing each Cyg~A measurement with its closest-in-time Cas~A, Tau~A, or Vir~A measurement.  We then measured the mean and the uncertainty in the mean for each of these (six; see Table~\ref{tab3}) collections of ratios, after rejecting outliers using Robust Chauvenet Rejection (RCR; \citealt{m17}).  More outliers were found with the Tau~A and Vir~A ratios than with the Cas~A ratios, because, being on the other side of the sky than Cyg~A, these pairs were generally observed with a greater separation in time.  The 2017 Cas~A ratios, however, were an exception, in that observations of Cyg~A began a full day before observations of Cas~A.

\begin{table*}
\caption{GBO 20-meter L- and X-band flux-density ratio measurements, with respect to Cyg~A.}\label{tab3}
\centering
\begin{tabular}{c c c c c c}
\hline 
\hline 
Mean Epoch & Band & Source & Number of Ratios & Number Rejected & $R_m$ \\
\hline 
\hline 
2014.53 & L & Cas~A & 48 & 3  & $1.123\pm0.019$	\\
2014.58 & X & Cas~A & 61 & 0  & $2.46\pm0.19$ \\
2016.74 & L & Cas~A & 43 & 1  & $1.115\pm0.021$	\\
2017.02 & L & Cas~A & 48 & 16 & $1.103\pm0.021$ \\
2017.02 & L & Tau~A & 48 & 11 & $0.529\pm0.012$ \\
2017.02 & L & Vir~A & 49 & 17 & $0.1370\pm0.0015$ \\
\hline 
\end{tabular}
\end{table*}

\section{GBO Models and Fits}\label{gbomodel}

\begin{figure}
\begin{center}
\includegraphics[width=0.45\textwidth]{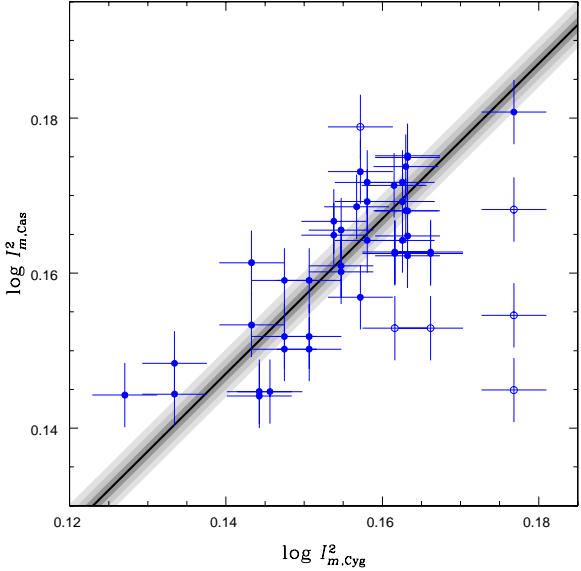}
\caption{Measured values of the square of the integral of the peak-normalized source function from Table 1, plotted $\log\Imcas$ vs. $\log\Imcyg$.  The line is the best fit of Equation~\ref{eq_Iab} to these, weighted data.  Open circles have been rejected using Robust Chauvenet Rejection (RCR; \citealt{m17}), and correspond to pairs of measurements that likely suffer from significantly different pointing errors; they do not contribute to the fit.  The shaded regions show the 1-, 2-, and 3-$\sigma$ statistical uncertainty in the best-fit value of $\log\beta$.  The uncapped error bars show the best-fit 1-$\sigma$ sample scatter, which in this case is two-dimensional (see \citealt{t11}, or \citealt{r01}).}
\label{fig_int2}
\end{center}
\end{figure} 

\begin{figure}
\begin{center}
\includegraphics[width=0.45\textwidth]{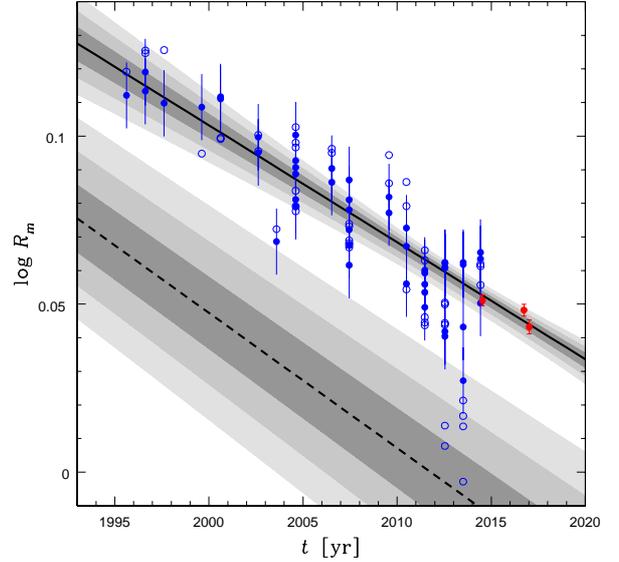}
\caption{23 years of GBO 40-foot (blue) and 20-meter (red) L-band Cas~A-to-Cyg~A flux-density ratios vs.\@ time, and best-fit model (solid line).  Open circles are measured values.  Filled circles are after subtracting off the best-fit third and fourth terms of Equation~\ref{eq_final}, which account for imperfect pointing and non-constant focus of the 40-foot telescope.  The shaded regions show the 1-, 2-, and 3-$\sigma$ statistical uncertainty in the best-fit values of $\log F_0$ and $m_{\nu_0}$.  The uncapped error bars show the best-fit 1-$\sigma$ sample scatter of the 40-foot measurements (that of the 20-meter measurements is almost negligible).  The significantly lower dashed line and shaded regions are from the fitted spectral and temporal models of \citet{b77}.}
\label{fig_erirafit}
\end{center}
\end{figure} 

\subsection{GBO 40-foot Pointing-Error and Non-Constant Focus Models} \label{correction}

From our first five years of GBO 40-foot observations, \citet{r00} found that the ratio of Cas~A's peak intensity to that of Cyg~A, $\Fmcas/\Fmcyg$, was significantly greater than expected, given the fitted spectral and temporal models of \citet{b77}.  In this section, we investigate if our 40-foot measurements could be biased high, due to (1) imperfect pointing of the telescope, and/or (2) non-constant focus of the telescope.
\\

\noindent
\textit{Imperfect Pointing:}  If during a drift scan the operator fails to zero-in on the true transit elevation of the source (\S\ref{gbo40data}), the peak intensity that we would then measure for that source would be underestimated.  Consequently, our higher-than-expected Cas~A-to-Cyg~A peak-intensity ratios could, at least in theory, be due to systematically greater pointing errors when we observed Cyg~A.  However, we regard this as unlikely:  Each year, our scans of Cyg~A were taken by the same, most-experienced member of our team, while most (but not all) of our scans of Cas~A were taken by first-time, student operators.  Hence, if anything, we would expect our Cas~A-to-Cyg~A peak-intensity ratios to be biased low due to imperfect pointing, not high.
\\

\noindent
\textit{Non-Constant Focus:}  Changes in temperature between observations of Cas~A and Cyg~A, and/or elevation-dependent differences in how gravity pulls on the dish and the receiver and its support structure could, at least in theory, change the focus of the telescope.  Non-ideal focus would result in a broader point-spread function, and consequently, a lower peak intensity.  Hence, if the conditions under which we took our Cyg~A observations were such that focus was systematically poorer than for our Cas~A observations, this could also explain our higher-than-expected Cas~A-to-Cyg~A peak-intensity ratios.  But again, we regard this as unlikely:  At the time of year that we took our observations, between June and August, both Cyg~A and Cas~A transit the meridian at night, or early in the morning, within 3.4 hours of each other, when temperatures at GBO are fairly stable.  Furthermore, both sources transit the meridian at similar, high elevations ($87^\circ$ for Cyg~A and $69^\circ$ for Cas~A), so we do not expect gravitational distortion of the telescope to differ significantly either.
\\

\indent
Although we anticipate that neither of these effects, imperfect pointing and non-constant focus, can account for our higher-than-expected Cas~A-to-Cyg~A peak-intensity ratios, we still model both effects (1) to be sure, and (2) to ensure that our fits to these data (see \S\ref{40ftresults}, \S\ref{combine}) are as accurate as possible.  As we will see in these sections, these effects are indeed small, generally offsetting, and insufficient to explain the discrepancy with \citet{b77}.
\\

\noindent
\textit{Pointing-Error Model}:  Imperfect pointing has two effects:  (1) The measured peak intensity is less than the true peak intensity ($F_m < F_t$); and (2) The measured peak-normalized source function is likely different from the true, or perfect-pointing, peak-normalized source function, resulting in $I_m^2 \neq I_t^2$.\footnote{In the idealized case of a Gaussian point-spread function, and a point, or Gaussian, source, $I_m^2 = I_t^2$, regardless of pointing.}  We model the consequent correlation, between measured peak intensity and the square of the integral of the measured peak-normalized source function, to first order, since pointing errors, if not negligible, were at least small (\S\ref{gbo40data}):
\begin{equation}\label{eq_fmft}
\frac{F_m}{F_t} = \left({\frac{I^2_m}{I^2_t}}\right)^\alpha \, ,
\end{equation}
\noindent where $\alpha$ is determined empirically in \S\ref{40ftresults} and \S\ref{combine}.\footnote{If the point-spread function has narrower wings than a Gaussian, which is usually the case, $I_m^2 < I_t^2$ and hence $\alpha$ must be positive; else, $\alpha < 0$.}     
\\

\noindent
\textit{Non-Constant Focus Model}:
Non-constant focus would imply that the 40-foot's true, or perfect-pointing, peak-normalized source function differs between our observations of Cas~A and Cyg~A.\footnote{Note, for the 40-foot at L band, this function is not appreciably different from the telescope's beam function, for either of these sources.  Cas~A is $\approx$6$\arcmin$ across and Cyg~A is $\approx$2$\arcmin$ across.  Convolution of the telescope's beam function with such, intrinsic distributions can be, at least roughly, approximated by adding their FWHMs in quadrature, which yields functions that are, at most, $\approx$0.3\% and $\approx$0.04\% wider, respectively.}  If there is a systematic component to this, we take it to be small, and, as above, model it to first order:\footnote{We also tried two-parameter models, such as $\Itcas=\beta\Itcyg+\gamma$ and $\Itcas=\beta I_{t,\mathrm{Cyg}}^{2\gamma}$, but the additional parameter was never requested by the data.}
\begin{equation}\label{eq_Iab}
\Itcas=\beta\Itcyg \, ,
\end{equation} 
where $\beta$ is also determined empirically in \S\ref{40ftresults} and \S\ref{combine}.  That said, we can place a prior on $\beta$, by fitting Equation~\ref{eq_Iab} to our measured values, $\Imcas$ and $\Imcyg$.\footnote{Specifically, we fit $\log\Itcas = \log\beta + \log\Itcyg$ to the logarithm of our measured values, while simultaneously rejecting outliers using Robust Chauvenet Rejection (RCR; \citealt{m17}), to avoid the fit being biased by pairs of measurements with significantly different pointing errors.  Indeed, $\Imcas < \Imcyg$ for five of the six rejected points (see Figure~\ref{fig_int2}), corresponding to some of our student-led observations of Cas~A having significantly greater pointing errors than their corresponding professional-led observations of Cyg~A, assuming $\alpha>0$ (see \S\ref{40ftresults} and \S\ref{combine}).  We measure the best-fit value of $\log\beta$, the statistical uncertainty in this best-fit value, and the sample scatter.  By the latter, we mean the scatter of the data about the model that is not accounted for by the data's error bars (which in this case, we take to be zero).  Like $\log\beta$, sample scatter is a model parameter, which we fit to the data (see \citealt{t11}, or \citealt{r01}, for a description of the maximum-likelihood technique).}  As can be seen in Figure~\ref{fig_int2}, these values appear to have changed in concert from year to year.  We find that that $\log\beta = 0.0070 \pm 0.0010$,\footnote{Note, if due only to the difference in these sources' intrinsic angular sizes, given the telescope's large beamwidth, $\log\beta$ would be no more than $\approx$0.003, and probably half that, both of which are ruled out.} with a best-fit sample scatter of 0.0059~dex.  Consequently, in our upcoming fits to 40-foot data (see \S\ref{40ftresults} and \S\ref{combine}), we adopt a Gaussian prior in $\log\beta$, of mean 0.0070 and width 0.0010~dex.

\begin{figure*}
\begin{center}
\includegraphics[width=\textwidth]{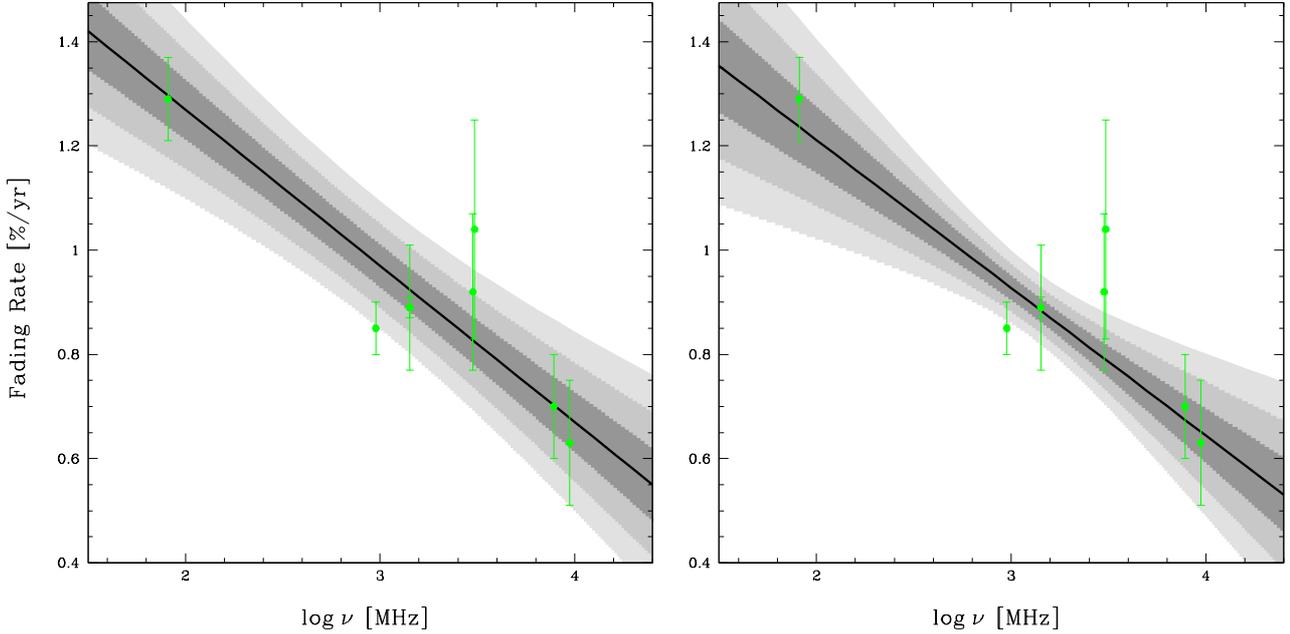}
\caption{Fading-rate measurements from Table~1 of \citet{b77}, and best-fit models.  The original result of \citet{b77} is on the left; ours is on the right.  The shaded regions show the 1-, 2- and 3-$\sigma$ statistical uncertainty in the best-fit values of $m_{\nu_0}$ and $m_{\Delta\log\nu}$.  The best-fit sample scatter (at least in the case of our fit) is negligible.  We were unable to reproduce the original result specifically, but do so generally.}
\label{fig-temp}
\end{center}
\end{figure*}

\begin{figure*}
\begin{center}
\includegraphics[width=\textwidth]{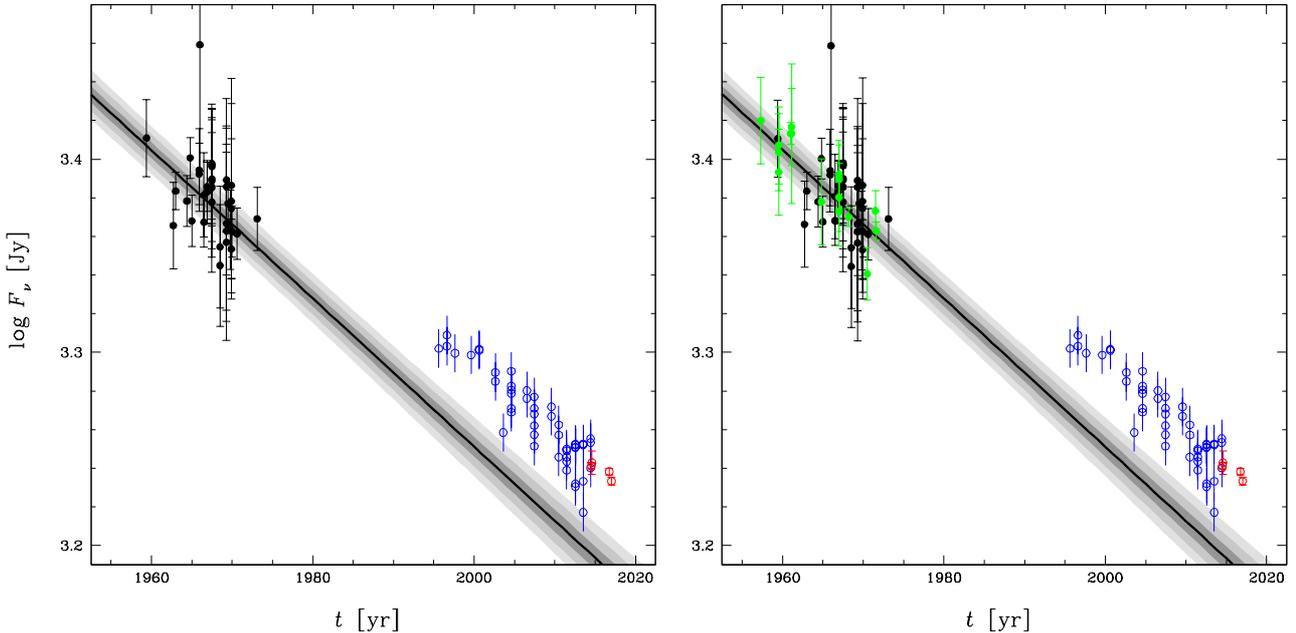}
\caption{Absolute (black) and relative (green) measurements of the flux density of Cas~A from \citet{b77}, and best-fit temporal models.  Relative measurements have been multiplied by the best-fit spectral model of the reference source, and all measurements have been referenced to 1405~MHz using the best-fit spectral model of Cas~A, for better visualization.  Open circles are the GBO 40-foot (blue) and 20-meter (red) measurements from Figure~\ref{fig-ball}, and are not included in these fits.  The fit on the left is to just the plotted, absolute Cas~A data.  The fit on the right is a simultaneous fit to all \citet{b77} flux-density data.  Both fits are additionally constrained by the fading-rate priors of \S\ref{btemp}, from the right panel of Figure~\ref{fig-temp}.  The shaded regions show the 1-, 2-, and 3-$\sigma$ statistical uncertainty in the best-fit values of $\log F_0$, $a_1$, $a_2$, $a_3$, $m_{\nu_0}$, and $m_{\Delta\log\nu}$.  The best-fit sample scatter is negligible.  This confirms that the \citet{b77} data do indeed imply a trend that undercuts our GBO data at later times (Figure~\ref{fig_erirafit}).}
\label{fig-bcas}
\end{center}
\end{figure*}

\begin{figure}
\begin{center}
\includegraphics[width=0.45\textwidth]{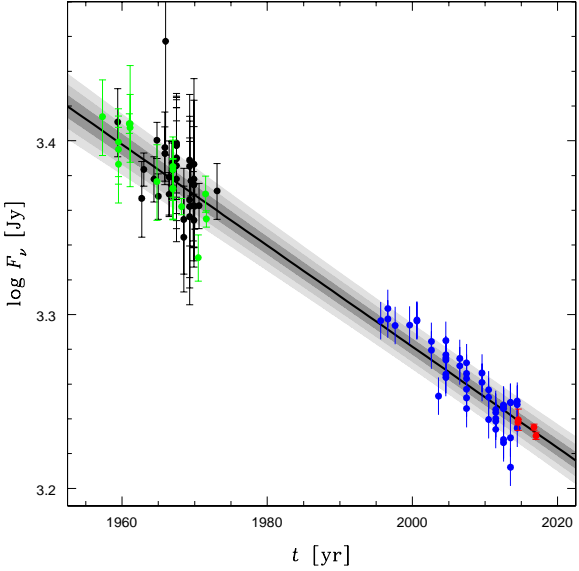}
\caption{Absolute (black) and relative (green) measurements of the flux density of Cas~A from \citet{b77}, our (also relative) GBO 40-foot (blue) and 20-meter (red) measurements, and our best-fit, one-segment temporal model of both epochs.  Relative measurements have been multiplied by the best-fit spectral model of the reference source, and all measurements have been referenced to 1405~MHz using the best-fit spectral model of Cas~A, for better visualization.  40-foot measurements have additionally been corrected for imperfect pointing and non-constant focus using the best-fit values of $\alpha$ and $\beta$, as in Figure~\ref{fig_erirafit}, also for better visualization.  The fading-rate priors of \S\ref{btemp} apply only to the epoch of the \citet{b77} data and consequently have not been included in this fit.  The shaded regions show the 1-, 2-, and 3-$\sigma$ statistical uncertainty in the best-fit values of $\log F_0$, $a_1$, $a_2$, $a_3$, $m_{\nu_0}$, and $m_{\Delta\log\nu}$.  The uncapped error bars show the best-fit 1-$\sigma$ sample scatter of the 40-foot measurements (that of the other measurements is negligible).}
\label{fig-single}
\end{center}
\end{figure}

\begin{figure*}
\begin{center}
\includegraphics[width=\textwidth]{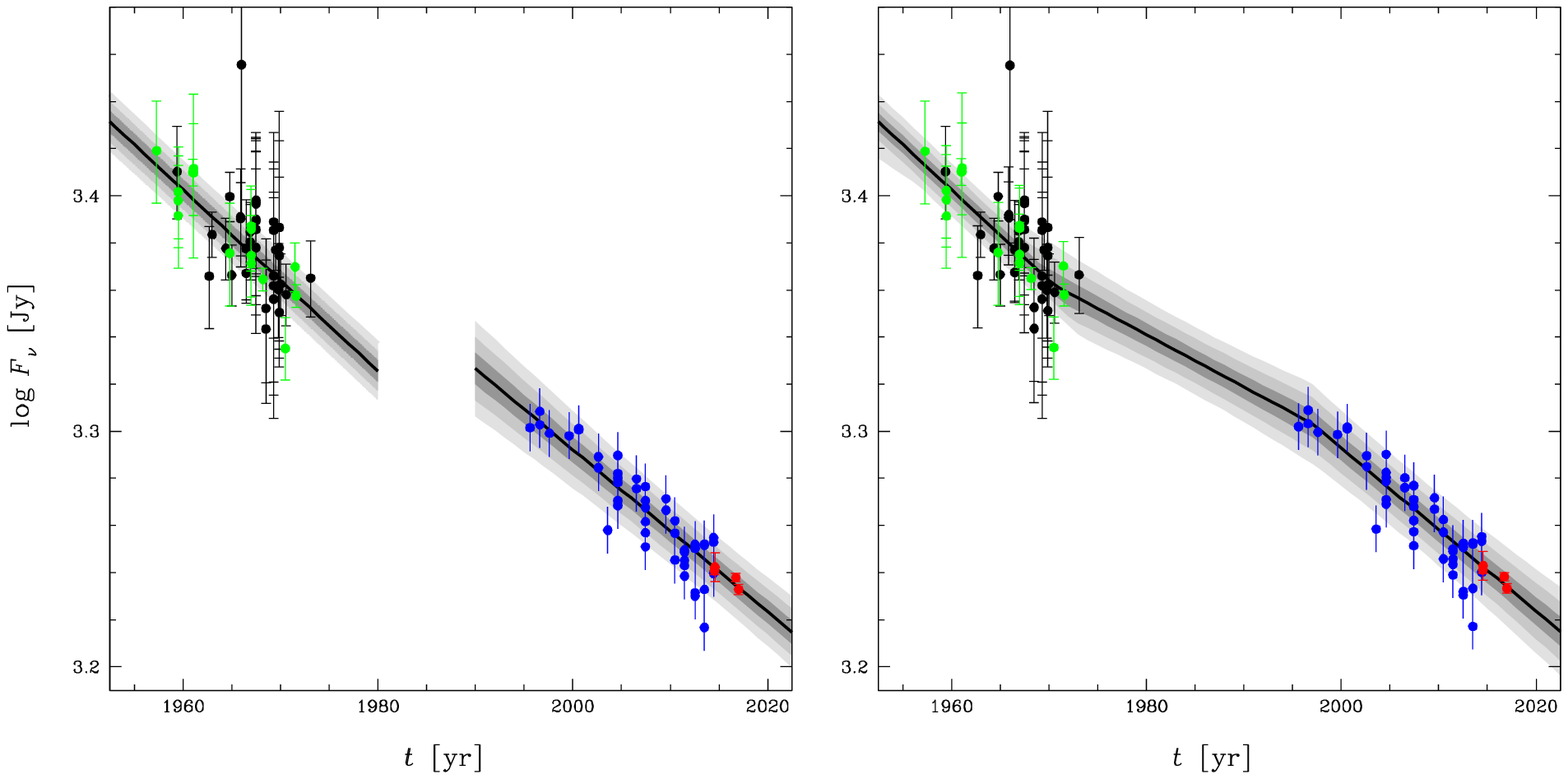}
\caption{The same as Figure~\ref{fig-single}, but for our best-fit two-segment (left) and three-segment (right) temporal models.  The fading-rate priors of \S\ref{btemp} have been applied to the first segment of each model.}
\label{fig-ball}
\end{center}
\end{figure*}

\subsection{GBO 40-foot and 20-meter Model and Fit} \label{40ftresults}

Of course, it is not the ratio of Cas~A's and Cyg~A's \textit{peak} intensities that we must model, but the ratio of their \textit{integrated} intensities, which do not depend on the telescope's precise pointing or focus, nor on the intrinsic size of the source.  With the 20-meter, this is straightforward:  We densely sample each source's 2D distribution on the sky, and can reconstruct and integrate these distributions (\S\ref{gbo20data}).  But with the 40-foot, we have only single, 1D slices through these 2D distributions (\S\ref{gbo40data}).  However, if one assumes radial symmetry, or at least the same shape, if not the same size, for these sources' 2D distributions, integrated intensity is then simply:
\begin{equation}
S_t \propto F_tI_t^2 \, ,
\end{equation}
\noindent
where subscript $t$ again refers to true, or perfect-pointing, values.  Given this, we define the measured Cas~A-to-Cyg~A flux-density ratio analogously:
\begin{equation}\label{eq_Rm}
R_m = \frac{S_{m,\mathrm{Cas}}}{S_{m,\mathrm{Cyg}}} \equiv \frac{\Fmcas\Imcas}{\Fmcyg\Imcyg} \, .
\end{equation}
\noindent These are the values that we list in Table~\ref{tab1}.  Note, Equation~\ref{eq_Rm} reduces to $R_m = F_{m,\mathrm{Cas}}/F_{m,\mathrm{Cyg}}$ in the limit of perfect pointing and identical 2D distributions.  
 
Substituting our pointing-error (Equation~\ref{eq_fmft}) and non-constant focus (Equation~\ref{eq_Iab}) models into Equation~\ref{eq_Rm} yields:
\begin{equation}\label{eq_meh1}
R_m = \frac{\Ftcas\Itcas}{\Ftcyg\Itcyg}\left(\frac{\Imcas}{\beta\Imcyg}\right)^{\alpha+1} 
    \equiv R_t\left(\frac{\Imcas}{\beta\Imcyg}\right)^{\alpha+1} \, ,
\end{equation}
where $R_t\equiv\Ftcas\Itcas/\Ftcyg\Itcyg$ is the true, or perfect-pointing, Cas~A-to-Cyg~A flux-density ratio at the time of the observations.

Given the size of Cyg~A's emitting region,\footnote{Cyg~A's lobes dominate the emission, with its nucleus contributing only $\approx$1~Jy at GHz frequencies (e.g., \citealt{k81}).} Cyg~A's brightness should not change over the timescale of these observations.  Consequently, the observed decrease in $R_t$ with time must be due to Cas~A fading.  In keeping with \citet{b77}, we model this fading exponentially:\footnote{\citet{v06} (also, see \citealt{s60}) argues that a power-law model would be more appropriate, physically, but also points out that this matters little, given the timescale over which these observations were made, compared to the age of Cas~A.}
\begin{equation} \label{eq_Rmt}
R_t(t) = R_0 e^{m_{\nu_0}(t-t_0)} \, ,
\end{equation}
where $R_0 \equiv R_t(t_0)$; $t_0$ is a reference time, which in this paper we set equal to the weighted mean of whichever observations we are fitting to;\footnote{Choice of reference value, be it a time or a frequency, affects the best-fit value of the normalization parameter, but not the best-fit model itself.  Given this freedom, one often selects simple, easy-to-use or easy-to-remember reference values; e.g., \citet{b77} used 1965 and 1~MHz, respectively.  However, choice of reference value also affects the degree to which the uncertainties in the best-fit parameter values are correlated (e.g., the uncertainties that \citealt{b77} measured are highly correlated).  By instead calculating reference values as we do here, we minimize these correlations (e.g., \citealt{m17}), the consequence being that we can easily, and analytically, approximate the uncertainty in our best-fit models, as a function of time in this case, and as a function of time and frequency in later cases (e.g., see Equation~\ref{envelope}).} and $m_{\nu_0}$ is the fading rate, in this case at $\nu_0 = 1405$~MHz.\footnote{In \S\ref{baars} and \S\ref{combine}, we instead use a frequency-dependent model.}  In units of percent per year, this is $-100 \times \ln 10 \times m_{\nu_0}$.  

Substituting Equation~\ref{eq_Rmt} into Equation~\ref{eq_meh1} and taking the logarithm of both sides yields:
\begin{equation} \label{eq_final}
\begin{split}
\log R_m(t) = &\log R_0 + m_{\nu_0}(t-t_0) \\ 
& +(\alpha+1)\log\left({\frac{\Imcas}{\Imcyg}}\right)-(\alpha+1)\log\beta  \, ,
\end{split}
\end{equation}
where the first two terms model the normalization and fading of the flux-density ratio, respectively; the third term accounts for imperfect pointing of the telescope, for 40-foot data; and the fourth term accounts for systematic differences in focus between observations of Cas~A and Cyg~A, again for 40-foot data.\footnote{The fourth term can also be used to account for systematic differences in the shape of each source's 2D distribution, but only in combination with data for which this does not need to be accounted for, such as our 20-meter data; see below.}  Equation~\ref{eq_final} can also, and simultaneously, be fitted to our 20-meter L-band Cas~A-to-Cyg~A flux-density ratios (Table~\ref{tab3}), by dropping the last two terms for these data.\footnote{Note, since the 40-foot and these 20-meter measurements were made at slightly different frequencies, 1405~MHz vs.\@ 1395~MHz, there should be a slight offset between their best-fit models.  However, given the spectral indices of Cas~A and Cyg~A at L band (see \citealt{b77}, \S\ref{ball}, and \S\ref{combine}), this offset should be only $\approx$7\% of the size of these 20-meter measurements' error bars (Table~\ref{tab3}), and consequently can be safely ignored.}

We plot our best fit of Equation~\ref{eq_final} to all 23 years of GBO 40-foot and 20-meter L-band measurements in Figure~\ref{fig_erirafit}.  We find that $\log R_0 = 0.0557^{+0.0011}_{-0.0019}$ (at $t_0 \equiv 2013.65$), $m_{\nu_0} = -0.00348^{+0.00022}_{-0.00023}$ (corresponding to $0.802^{+0.053}_{-0.050}$~\%/yr), $\alpha = 0.64 \pm 0.15$, and $\log\beta = 0.00754^{+0.00090}_{-0.00089}$, with best-fit sample scatters of 0.0099~dex for the 40-foot data and 0.0014~dex for the 20-meter data.    

The fitted pointing-error and non-constant focus ``correction'' terms in Equation~\ref{eq_final} are, as expected, small, but not negligible.  The mean pointing-error correction is $0.00774 \pm 0.00073$~dex, or $1.80 \pm 0.17$~\%.  This suggests that, on average, our pointing was a bit better for Cyg~A than for Cas~A, as expected (\S\ref{correction}).  The non-constant focus correction is $-0.0123 \pm 0.0019$~dex, or $-2.80 \pm 0.42$~\% (this uncertainty is 44\% anti-correlated with the uncertainty in the mean pointing-error correction).  This suggests that the 40-foot's focus was a bit worse when observing Cas~A than when observing Cyg~A, for either of the reasons postulated in \S\ref{correction}.  The former correction, visibly, accounts for some of the scatter of the data about the best-fit model (Figure~\ref{fig_erirafit}), and raises the best-fit model.  The latter correction lowers the best-fit model, to be in accordance with the 20-meter data, although $\log\beta$ is also constrained by the prior that we were able to place on it in \S\ref{correction}.  The sum of these corrections is $-0.0046 \pm 0.0015$~dex, or $-1.05 \pm 0.35$~\%, which falls significantly short of the roughly $-0.06 \pm 0.01$~dex, or $-15 \pm 2$~\% correction that would be required to explain the discrepancy with the fitted spectral and temporal models of \citet{b77}, first identified by \citet{r00} (Figure~\ref{fig_erirafit}).  Furthermore, the 20-meter measurements, which are not subject to these corrections, confirm this discrepancy.  

\section{Models and Fits to the Baars et al.\@ (1977) Data}\label{baars}

Now that we have established that Cas~A is, and has been since at least the mid-1990s, significantly brighter than expected, given the fitted spectral and temporal models of \citet{b77}, at least in L band, two likely explanations remain:  (1) Cas~A rebrightened sometime between the end of the \citet{b77} observations and the beginning of our GBO observations, so between the mid-1970s and mid-1990s, after which it resumed fading; or (2) the fitted spectral and/or temporal models of \citet{b77} are erroneous.  Given this, and given the importance of these fitted models for flux-density calibration in radio astronomy, we now re-analyze the original \citet{b77} data.  

\subsection{Temporal Model and Fit} \label{btemp}

We begin by fitting the same, frequency-dependent fading-rate model to the same fading-rate measurements that \citet{b77} did.  The fading-rate model is given by:
\begin{equation}\label{temp}
-\frac{100}{F_\nu}\frac{dF_\nu}{dt} = m_{\nu_0}+m_{\Delta\log\nu}\log\left(\frac{\nu}{\nu_0}\right) \, ,
\end{equation}
where we have decided to normalize the function at the weighted mean of the data's $\log\nu$ values (corresponding to $\nu_0 \equiv 1315$~MHz), instead of at 1~GHz.  The fading-rate measurements are from Table~1 of \citet{b77}.

We plot our best fit of Equation~\ref{temp} to these measurements in the right panel of Figure~\ref{fig-temp}.  We find that $m_{\nu_0} = 0.894 \pm 0.021$~\%/yr and $m_{\Delta\log\nu} = -0.284 \pm 0.053$~\%/yr, with a best-fit sample scatter of zero.  This differs from the original result of \citet{b77} (left panel), which has marginally different best-fit values, but significantly different uncertainties in these best-fit values:  \citet{b77} appears to have overstated the uncertainty in the best-fit model around 1~GHz, and understated it below $\approx$100~MHz.\footnote{In an attempt to determine what \citet{b77} might have done incorrectly, or at least differently, we also tried an unweighted fit, with $\nu_0 \equiv 1$~GHz (as in Equation~\ref{eq1}).  Although this was much closer to the original result, it still did not reproduce it exactly.}  Despite this, our fitted temporal models are sufficiently similar that this is unlikely to be the source of the discrepancy in Figure~\ref{fig_erirafit}.

\subsection{Spectral Model and Fits} \label{ball}

\citet{b77} used their best-fit temporal model to reference all of their measurements to epoch 1965, after which they fitted their spectral model to these ``corrected'' measurements.  Generally, this is a poor modeling practice, in that it is seldom clear how all of the various uncertainties propagate.\footnote{If the measurement uncertainties are small, and the uncertainties in the best-fit values of $m_{\nu_0}$ and $m_{\Delta\log\nu}$ are also small, and uncorrelated, this can be done with some reliability -- for a single measurement.  But for multiple measurements, this only introduces correlations between the ``corrected'' measurements, which is problematic when one attempts to fit a model to them, because most statistics assume fully independent data.  Alternatively, if the uncertainties in the best-fit values of $m_{\nu_0}$ and $m_{\Delta\log\nu}$ are simply ignored, as was likely the case here, each ``corrected'' measurement will be in error by some, different amount, and these errors are not guaranteed to be offsetting.  In short, it is always better to ``bring the model to the data'' (e.g., as we did in \S\ref{gbomodel}), than to attempt the opposite.  This said, note that one often references data to a common frequency, or epoch, etc., when plotting.  But this is for visualization only.  Model fits to these data should always be ``forward folded''.}  Instead, we fit our spectral and temporal models simultaneously (\S\ref{historical}), adopting the results of the previous section as priors.  Otherwise, our only difference is that we use a slightly different spectral model:  Where \citet{b77} used piecewise linear and quadratic models, we use a third-order polynomial model over the entire, 22~MHz to 22~GHz, spectral range.  Our combined spectral and temporal model is then given by:
\begin{equation}\label{model}
\begin{split}
\log F_m(\nu,t) = &\log F_0 + a_1\log\left(\frac{\nu}{\nu_\mathrm{Cas}}\right) \\
& + a_2\left[\log\left(\frac{\nu}{\nu_\mathrm{Cas}}\right)\right]^2  
+ a_3\left[\log\left(\frac{\nu}{\nu_\mathrm{Cas}}\right)\right]^3 \\
& + \left[m_{\nu_0}+m_{\Delta\log\nu}\log\left(\frac{\nu}{\nu_0}\right)\right](t-t_0) \, ,
\end{split}  
\end{equation}
\noindent
where the first term is the normalization; the next three terms are our spectral model, normalized at the weighted mean of the data's $\log\nu$ values; and the last term is our frequency-dependent temporal model, where, \textit{a priori}, we take $m_{\nu_0}$ to be distributed as a Gaussian of mean 0.894~\%/yr and width 0.021~\%/yr, $m_{\Delta\log\nu}$ to be distributed as a Gaussian of mean -0.284~\%/yr and width 0.053~\%/yr, and $\nu_0 \equiv 1315$~MHz (\S\ref{btemp}). 

\begin{table*}
\caption{Best-fit parameter values and 1-$\sigma$ statistical uncertainties in these values for our favored, three-segment model.}\label{tab4}
\centering


\begin{tabular}{c c c c}
\multicolumn{4}{c}{\textbf{Cas~A Segment Specific}}\\
\hline 
\hline 
Parameter & Segment 1 & Segment 2 & Segment 3 \\
\hline 
\hline 
$\log F_0$ & $3.3731 \pm 0.0037$ & $3.3173 \pm 0.0047$ & $3.2530^{+0.0051}_{-0.0050}$ \\
\\
$m_{\nu_0}$ & $-0.003884^{+0.000092}_{-0.000093}$ & $-0.00227 \pm 0.00024$ & $-0.00350^{+0.00022}_{-0.00025}$ \\
\\
& ($0.894 \pm 0.021$~\%/yr) & ($0.522^{+0.056}_{-0.054}$~\%/yr) & ($0.806^{+0.058}_{-0.050}$~\%/yr) \\
\\
reference time & 1963.9 & 1983.8 & 2006.9 \\
\end{tabular} \\

\bigskip

\begin{tabular}{c c c c c }
\multicolumn{5}{c}{\textbf{Source Specific}} \\
\hline 
\hline 
Parameter & Cas~A & Cyg~A & Tau~A & Vir~A \\
\hline 
\hline 
$\log F_0$ & segment specific & $3.1861^{+0.0046}_{-0.0047}$ & $2.9083^{+0.0044}_{-0.0045}$ & $2.3070^{+0.0045}_{-0.0046}$ \\
\\
$a_1$ & $-0.732^{+0.011}_{-0.010}$ & $-1.038\pm0.011$ & $-0.226^{+0.014}_{-0.013}$ & $-0.876\pm0.017$  \\
\\
$a_2$ & $-0.0094^{+0.0058}_{-0.0059}$ & $-0.1457^{+0.0075}_{-0.0077}$ & $-0.0113^{+0.0081}_{-0.0083}$ & $-0.047^{+0.031}_{-0.033}$ \\
\\
$a_3$ & $0.0053\pm0.0058$ & $0.0170^{+0.0075}_{-0.0076}$ & $-0.0275^{+0.0077}_{-0.0079}$ & $0.073^{+0.030}_{-0.029}$ \\
\\
reference frequency & 1477~MHz & 1416~MHz & 1569~MHz & 1482~MHz \\
\hline 
$m_{\nu_0}$ & segment specific & -- & $-0.00044^{+0.00019}_{-0.00018}$ & -- \\
\\
& & & ($0.102^{+0.042}_{-0.043}$~\%/yr) & \\
\\
reference time & segment specific & -- & 2009.05 & -- \\
\hline 
$m_{\Delta\log\nu}$ & $0.00124 \pm 0.00018$ & -- & -- & -- \\
\\
& ($-0.286\pm0.042$~\%/yr) & & & \\
\\
reference frequency & 1315~MHz & -- & -- & -- \\
\\
reference time & 2005.64 & -- & -- & -- \\
\hline 
\end{tabular}\\
\bigskip

\begin{tabular}{c c c}
\multicolumn{3}{c}{\textbf{Telescope Specific}} \\
\hline 
\hline 
Parameter & 40-foot & All Others \\
\hline 
\hline 
$\alpha$ & $0.64^{+0.16}_{-0.15}$ & -- \\
\\
$\beta$ & $0.00757^{+0.00091}_{-0.00089}$ & -- \\
\\
scatter & $0.0099$ & $0$  \\
\hline 
\end{tabular}
\end{table*}

In the left panel of Figure~\ref{fig-bcas}, we fit Equation~\ref{model} to the exact same measurements that \citet{b77} did, the absolute measurements of Cas~A listed in their Table~2.  In the right panel, we expand upon this by additionally fitting to all of their relative measurements of Cas~A, which are respect to Vir~A.  This requires simultaneously fitting to Vir~A, and as some of these measurements are relative to Cyg~A and Tau~A, this requires simultaneously fitting to all four sources.  For this, we adopt the same, third-order polynomial spectral model for each source, but with independent parameters and normalization reference frequencies.\footnote{Altogether, this is a 22-parameter model, but not a problem for our \textit{Galapagos} software (\S\ref{gbo40data}).}  In either case, we confirm that the \citet{b77} data do indeed imply a trend that undercuts our GBO data at later times, albeit by a marginally smaller amount, but also at a greater level of significance, compared to the original result (Figure~\ref{fig_erirafit}). 

\section{Combined Models and Fits}\label{combine}

We have now established that neither our more recent, GBO measurements (\S\ref{gbomodel}) nor the original, fitted spectral and temporal models of \citet{b77} (\S\ref{baars}) have misdirected us (although we were not able to reproduce the latter exactly, and with more powerful statistical and computing techniques, have been able to improve on these models, especially within certain frequency ranges; see below).  Consequently, we conclude that Cas~A rebrightened, or at least faded at a much slower rate, between the end of the \citet{b77} observations and the beginning of our GBO observations (\S\ref{baars}).

In this section, we begin by modeling the average fading rate of Cas~A between the late 1950s and late 2010s, by ignoring the fading-rate priors of \S\ref{btemp}, which clearly apply only to the epoch of the \citet{b77} data.  We then reintroduce these priors, and model the fading of Cas~A in two ways:  (1) assuming a rebrightening event between the last \citet{b77} observation and our first GBO observation; and (2) assuming a period of much slower fading between these epochs.

\begin{figure*}
\begin{center}
\includegraphics[width=\textwidth]{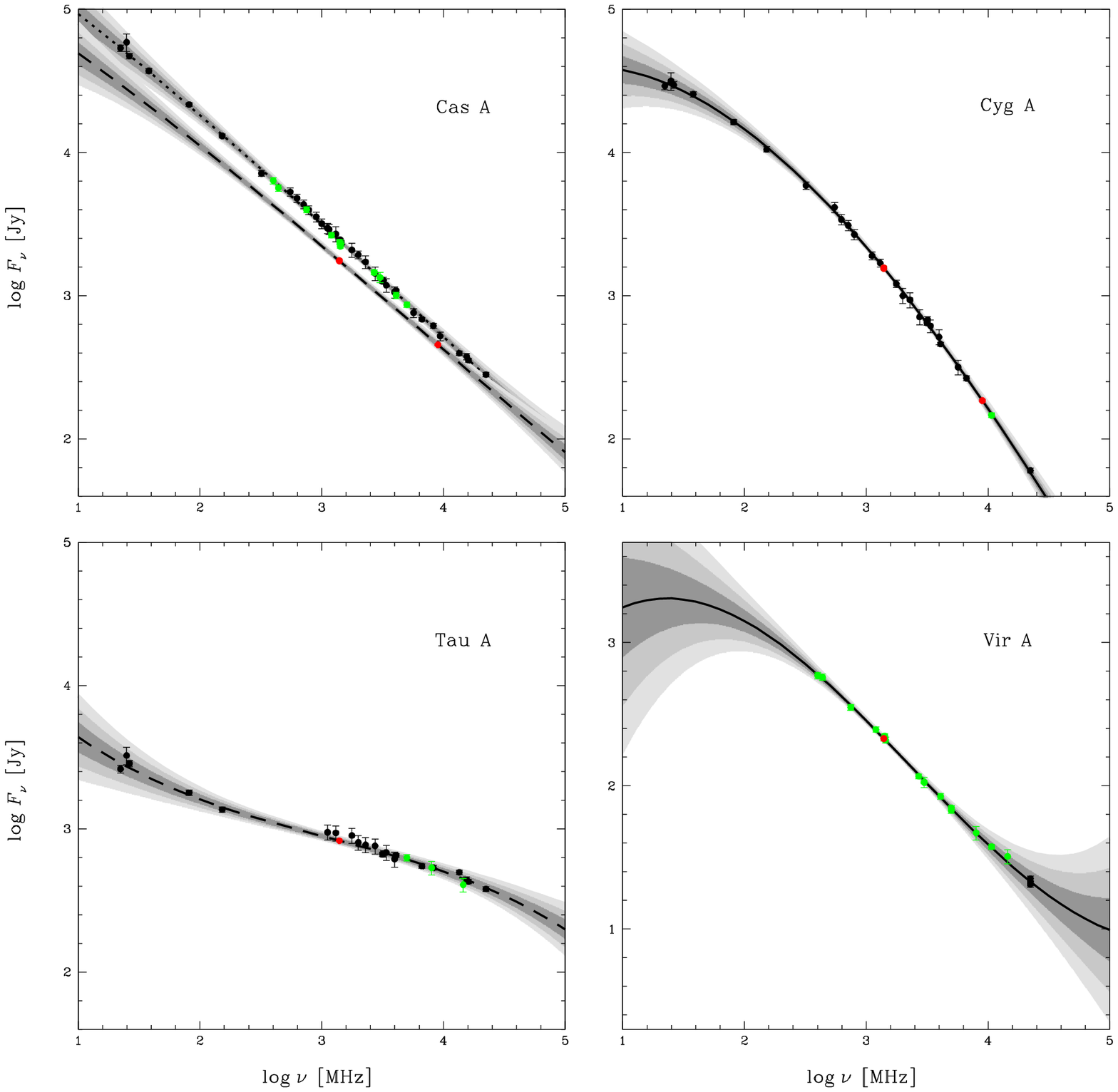}
\caption{Our best-fit spectral models (specifically, corresponding to the best-fit three-segment temporal model of Figure~\ref{fig-ball}).  The data are color-coded as in Figures \ref{fig-single} and \ref{fig-ball}.  Note, however, that we have not plotted the GBO 40-foot data here, for greater clarity at this frequency.  Relative measurements (green and red) have again been multiplied by the best-fit spectral model of the reference source, and (1) the \citet{b77} Cas~A measurements have been referenced to 1965 (dotted curve), (2) our GBO Cas~A measurements have been referenced to 2015 (dashed curve), and (3) all Tau~A measurements have been referenced to 2015 (dashed curve) using the best-fit temporal models of Cas~A and Tau~A, respectively, for better visualization.  The shaded regions show the 1-, 2-, and 3-$\sigma$ statistical uncertainty in the best-fit values of $\log F_0$, $a_1$, $a_2$, $a_3$, and, when applicable, $m_{\nu_0}$, and $m_{\Delta\log\nu}$.  The best-fit sample scatter of all but the 40-foot measurements is negligible.}
\label{fig-spec}
\end{center}
\end{figure*} 

\begin{figure*}
\begin{center}
\includegraphics[width=\textwidth]{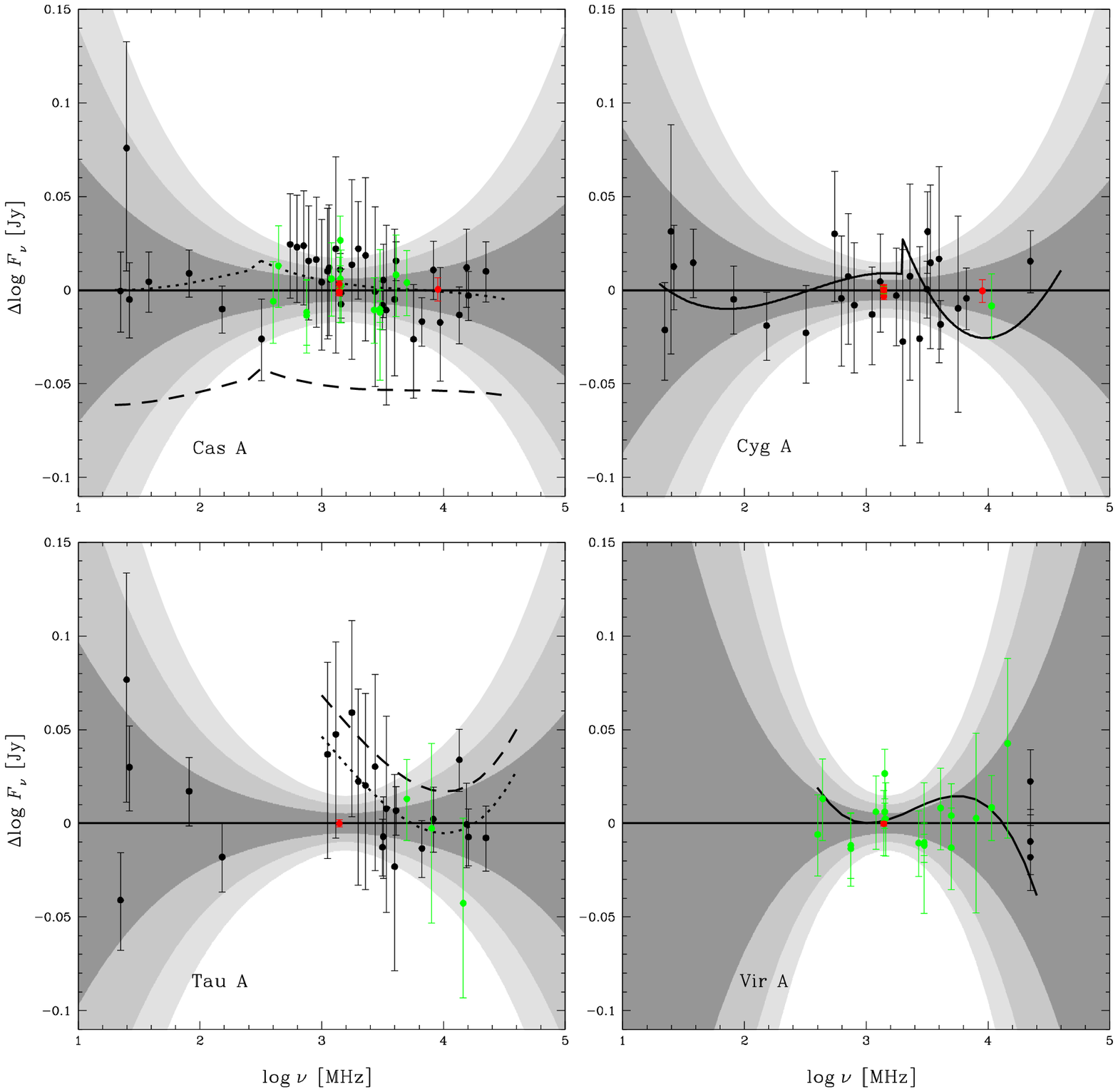}
\caption{Difference between the original, best-fit spectral models of \citet{b77} and our updated and improved best-fit spectral models (Figure~\ref{fig-spec}).  The \citet{b77} Cas~A and Tau~A models have been referenced to 1965 (dotted) and 2015 (dashed), respectively.  Our Cas~A and Tau~A models have been referenced to 2015.  Clearly, \citet{b77} overpredicted the fading of Cas~A, and appear to have also underpredicted (specifically, set to zero) the fading of Tau~A.  But even the 1965 versions of these curves differ by up to $\approx$6\% -- $\approx$11\% in the couple- to many-GHz range of some of these spectra (Cyg~A and Tau~A in particular).}
\label{fig-specdiff}
\end{center}
\end{figure*} 

\begin{figure}
\begin{center}
\includegraphics[width=0.45\textwidth]{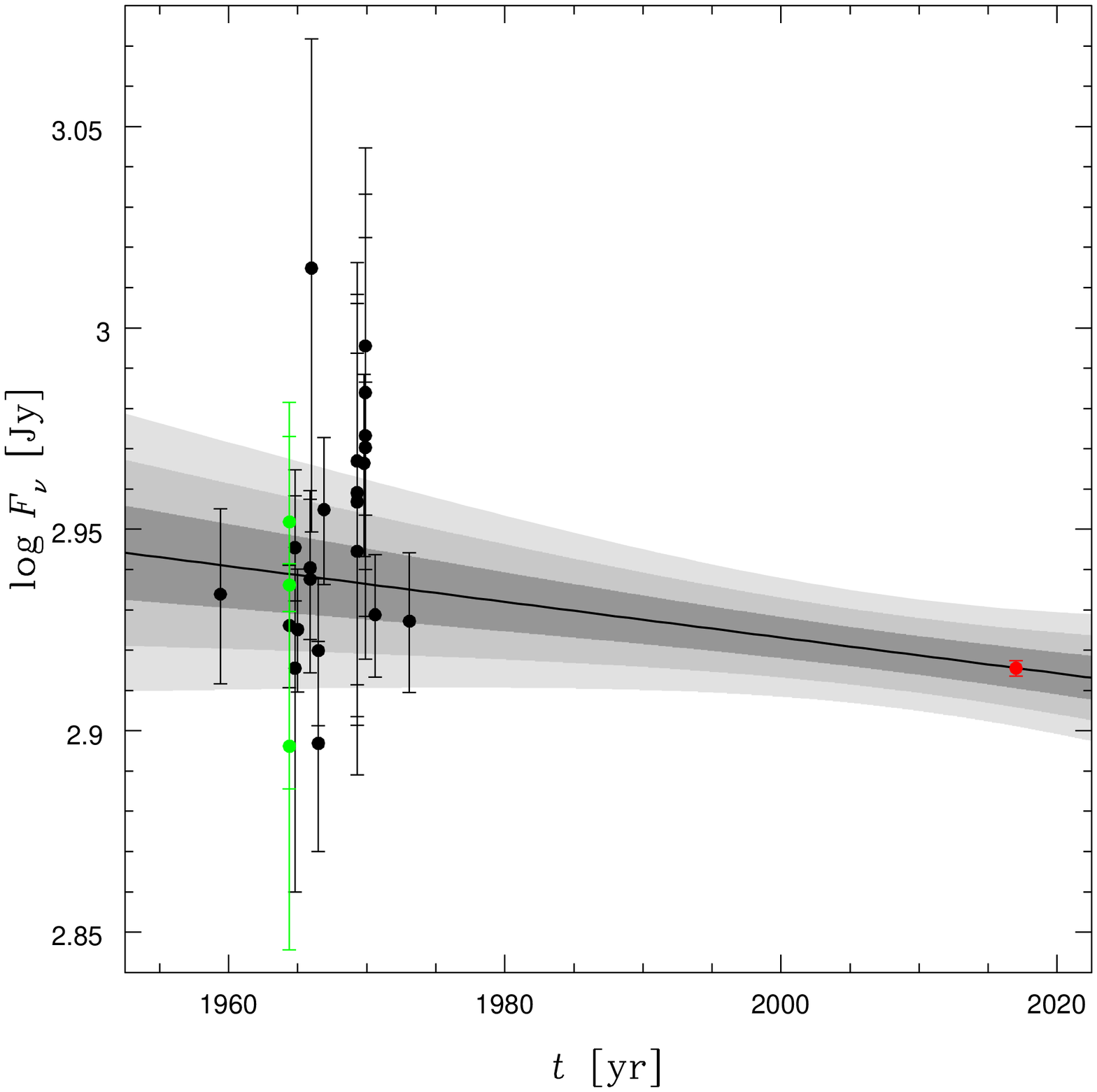}
\caption{The same as Figure~\ref{fig-single}, but for Tau~A.  The fading of Tau~A appears to be measurable over this timescale.}
\label{fig-tau}
\end{center}
\end{figure} 

For both of these models, we fit to all data (Cas~A, Cyg~A, Tau~A, and Vir~A) simultaneously, as we did in \S\ref{ball}.  Given the longer timescale over which these observations were made, we also permit Tau~A to fade (neither Cyg~A nor Vir~A should).  However, we do not have enough late-time data to measure if this fading also has a frequency dependence, so for Tau~A, we simply measure its fading rate in L band.  Furthermore, we do not have enough late-time data to measure if the frequency dependence of Cas~A's fading rate is also changing, so we take it to be constant across all epochs. 

\subsection{Average Fading Model and Fit}\label{average}

Our fading-rate priors (\S\ref{btemp}) derive from eight fading-rate measurements.  On average, these measurements were made over $12.9 \pm 4.4$ year spans of time, beginning in $1959.1 \pm 4.8$ and ending in $1972 \pm 2.1$.  As such, these priors are germane really only to this relatively brief period, which matches the epoch of the rest of the \citet{b77} data almost perfectly.  So, if we wish to characterize Cas~A with a single fading rate, representative of all of the epochs for which we have data, we should not necessarily bind ourselves with these priors.  Dropping them for this fit only yields:  $\log (F_0/1~\mathrm{Jy}) = 3.2494^{+0.0041}_{-0.0042}$ (at $t_0 = 2005.64$), $m_{\nu_0} = 0.663 \pm 0.019$~\%/yr, $m_{\Delta\log\nu} = -0.317 \pm 0.067$~\%/yr, and a best-fit sample scatter for all but the 40-foot data (i.e., for all data with measured uncertainties) of zero (see Figure~\ref{fig-single}).  At 1405~MHz, this corresponds to a fading rate of $0.670 \pm 0.019$~\%/yr, which is consistent with our initial finding of $0.62 \pm 0.12$~\%/yr (\citealt{r00}).  This is also consistent with the finding of \citet{v07a}, who measured a fading rate of $0.67\pm0.04$~\%/yr between 1977 and 2004 at the nearby frequency of 927~MHz using the 10-meter diameter radio telescope at Staraya Pustyn' Radio Astronomy Observatory in Russia.  However, all of these measurements are inconsistent with the fading rate measured during the epoch of the \citet{b77} data:  $0.93 \pm 0.04$~\%/yr at 1405~MHz as measured by \citet{b77}, or $0.886 \pm 0.021$~\%/yr at 1405~MHz as we have remeasured it, in \S\ref{btemp}.

\subsection{Rebrightening Models and Fits}\label{rebrightening}

Next, we fit two- and three-segment models to the same data.  In the case of the two-segment model, we fit one segment to the \citet{b77} data, one segment to our GBO data, and we assume that an, unmodeled, rebrightening event took place between these segments.  In the case of the three-segment model, we assume a less-dramatic rebrightening, consisting merely of a period of slower fading, between two faster-fading segments.  In both cases, we include the fading-rate priors of \S\ref{btemp}, applying them to the first segment only.  We plot our best-fit two- and three-segment models in the left and right panels of Figure~\ref{fig-ball}, respectively.  

Compared to a one-segment model with the fading-rate priors also applied, these two- and three-segment, rebrightening models are favored at the 6.8$\sigma$ ($\Delta\chi^2 = 49.8$, $\Delta\nu = 2$) and 6.3$\sigma$ ($\Delta\chi^2 = 50.4$, $\Delta\nu = 4$) credible levels, respectively.  

\citet{v06}, \citet{v07a}, and \citet{v14} (and references therein) observed Cas~A and Cyg~A between the two primary epochs considered in this paper, and at a variety of frequencies.  We did not include these data in our analysis, because the scatter of these data (compared to their measured uncertainties) suggests that systematics might be an issue, and we do not know enough about these telescopes to model possible systematics (e.g., as we did in \S\ref{gbomodel}).  However, these data are still useful to us in that they show no evidence of a sudden rebrightening, but are generally consistent with monotonic fading, which favors our three-segment model over our two-segment model.  We present best-fit parameter values and uncertainties for this, 27-parameter, model in Table~\ref{tab4}.  

Both of our rebrightening models, as well as the one-segment model from \S5.1, result in nearly identical fitted spectral models.  We plot these for our favored, three-segment model, and for all four sources, in Figure~\ref{fig-spec} (also, see \citealt{pb16} for an independent analysis).  

To examine the quality of the original, fitted spectral models of \citet{b77}, we plot their difference from our updated and improved versions in Figure~\ref{fig-specdiff}.  As we have already established, \citet{b77} overpredicted the fading of Cas~A, and consequently their Cas~A spectrum is now too low.  Conversely, they assumed no fading for Tau~A, and consequently their Tau~A spectrum now appears to be too high (see below).

However, even if one restricts oneself to an apples-to-apples comparison at epoch 1965, there are still significant differences.  In particular, the \citet{b77} Cyg~A spectrum is (1) discontinuous, (2) significantly high (up to $\approx$7\%) in the few-GHz range, and (3) significantly low (up to $\approx$6\%) in the many-GHz range.  And the \citet{b77} Tau~A spectrum is significantly high (up to $\approx$11\%) in the couple-GHz range.  These differences are much larger than the $\approx$2\% accuracy quoted by \citet{b77}.

Finally, for Tau~A, we measure a fading rate of $m_{\nu_0} = 0.102^{+0.042}_{-0.043}$~\%/yr in L band (see Figure~\ref{fig-tau}).  \citet{v07b} found a similar result:  $0.18\pm0.10$~\%/yr, from observations made between 1977 and 2000 at 927~MHz.  

For all of our fitted models, we have made every effort to select reasonable reference frequencies and times, in an attempt to minimize correlations between the fitted parameters.  Given this, one may approximate the positive and negative 1-$\sigma$ uncertainties in any of our best-fit spectral and temporal models, at any frequency and time, by taking the square root of the following function:
\begin{equation}
\begin{split}
\sigma_{\log F_m}^2(\nu,t) \approx\,\, &\sigma_{\log F_0}^2 + \sigma_{a_1}^2\left[\log\left(\frac{\nu}{\nu_\mathrm{source}}\right)\right]^2 \\
& + \sigma_{a_2}^2\left[\log\left(\frac{\nu}{\nu_\mathrm{source}}\right)\right]^4  
+ \sigma_{a_3}^2\left[\log\left(\frac{\nu}{\nu_\mathrm{source}}\right)\right]^6 \\
& + \sigma_{m_{\nu_0}}^2(t-t_{\mathrm{segment}})^2 \\
& +\sigma_{m_{\Delta\log\nu}}^2\left[\log\left(\frac{\nu}{\nu_0}\right)\right]^2(t-t_\mathrm{source})^2 + \sigma_\mathrm{scatter}^2 \, ,
\end{split} 
\label{envelope}
\end{equation}
\noindent where $\sigma_X$ is the positive or negative 1-$\sigma$ statistical uncertainty, respectively, in the best-fit value of parameter $X$; the two temporal terms are only for Cas~A and Tau~A, and in the case of Tau~A, the reference time $t_\mathrm{segment} = t_\mathrm{source}$; and for all four sources, the best-fit scatter $\sigma_\mathrm{scatter} = 0$ in all of these fits (e.g., Table~\ref{tab4}).

\section{Conclusions}\label{conclusion}

The fading of Cas~A appears to be more complicated than originally measured, with the remnant fading quickly between the late 1940s and late 1960s/early 1970s ($0.886 \pm 0.021$~\%/yr in L band), then more slowly until the mid- to late 1990s ($0.516^{+0.056}_{-0.054}$~\%/yr in L band), and then quickly again until at least the late 2010s ($0.798^{+0.058}_{-0.050}$~\%/yr in L band; \S\ref{rebrightening}).\footnote{It should be noted that Cas~A also appears to have flared, at the $\approx$4$\sigma$ confidence level, for a brief time in the 1970s, but only at low frequencies (38~MHz; \citealt{ep75}, \citealt{r77a}, \citealt{r77b}).}  Whether Cas~A will continue to fade at this rate, or over longer timescales will more closely adhere to the average that we measured between the late 1950s and late 2010s ($0.654 \pm 0.019$ \%/yr in L band; \S\ref{average}) is unknown.  Consequently, care should be taken if calibrating against Cas~A, or Tau~A for that matter, although Tau~A is fading much more slowly, and consequently should remain reasonable as a calibrator longer.  

Furthermore, we have identified potential weaknesses in the absolute spectral calibrations provided by \citet{b77}, particularly for Cyg~A and Tau~A in the couple- to many-GHz range, not due to measurement errors, but due, perhaps, to statistical and computing limitations at the time (\S\ref{rebrightening}).  We have updated and improved these calibrations, for Cas~A, Cyg~A, Tau~A, and Vir~A, and have provided a better way to characterize the uncertainty in these calibrations, as a function of frequency, and, in the case of Cas~A and Tau~A, as a function of time.

This work can be improved upon in the future with additional measurements of Cas~A at other frequencies, to see if the frequency dependence of its fading rate is also changing, and with additional measurements of Tau~A, both at L band and at other frequencies, to measure the frequency dependence of its fading rate.

\section*{Acknowledgements}

We gratefully acknowledge the support of the National Science Foundation, through the following programs and awards:  ESP 0943305, MRI-R$^2$ 0959447, AAG 1009052, 1211782, and 1517030, ISE 1223235, HBCU-UP 1238809, TUES 1245383, and STEM$+$C 1640131.  We are also appreciative to have been supported by the Mt.\@ Cuba Astronomical Foundation, the Robert Martin Ayers Sciences Fund, and the North Carolina Space Grant Consortium.  We also thank the referee, and the editor, for their helpful comments and suggestions.  Finally, we thank the approximately 400 students and educators who have participated in \textit{Educational Research in Radio Astronomy (ERIRA)} at Green Bank Observatory since 1992, many of whom helped to collect the 40-foot data that we present in this paper.  We also thank Green Bank Observatory for hosting our group for all of these years.

\end{document}